\documentclass[conference, a4paper]{IEEEtran}

\usepackage[switch]{lineno}

\usepackage{url}
\usepackage{acronym}
\usepackage{adjustbox}
\usepackage{amsmath,bm}
\usepackage{booktabs}
\usepackage{caption}
\usepackage{cite}
\usepackage{dirtytalk}
\usepackage{enumitem}
\usepackage{graphicx}
\usepackage{makeidx}
\usepackage{multirow}
\usepackage{multicol}
\usepackage[ruled]{algorithm2e}
\usepackage{subcaption}
\usepackage[table,xcdraw]{xcolor}
\usepackage{tabularx}
\usepackage{url}
\usepackage{soul}
\usepackage[group-separator={\,}]{siunitx}
\usepackage{comment}
\usepackage{tablefootnote}
\usepackage[table]{xcolor}
\usepackage{lipsum}
\usepackage[justification=centering]{caption}

\makeatletter
\newcommand{\algorithmfootnote}[2][\footnotesize]{%
  \let\old@algocf@finish\@algocf@finish
  \def\@algocf@finish{\old@algocf@finish
    \leavevmode\rlap{\begin{minipage}{\linewidth}
    #1#2
    \end{minipage}}%
  }%
}
\makeatother
\acrodef{ANOVA}	{Analysis of Variance}
\acrodef{HMD}	{Head Mounted Display}
\acrodef{PENS}	{Player Experience of Need Satisfaction}
\acrodef{QoE}	{Quality of Experience}
\acrodef{VE}	{Virtual Environment}
\acrodef{VR} 	{Virtual Reality}
\newif\ifcomment

\makeatletter
\patchcmd{\@maketitle}
  {\addvspace{0.5\baselineskip}\egroup}
  {\addvspace{-0.1\baselineskip}\egroup}
  {}
  {}
\makeatother

\begin{document}

\IEEEoverridecommandlockouts


\title{Towards Deep Learning Methods for Quality Assessment of Computer-Generated Imagery
}

\author{

   \IEEEauthorblockN{Markus Utke\IEEEauthorrefmark{1}, Saman Zadtootaghaj\IEEEauthorrefmark{1}, Steven Schmidt\IEEEauthorrefmark{1}, Sebastian M{\"o}ller\IEEEauthorrefmark{1}\IEEEauthorrefmark{2}}
    
 \IEEEauthorblockA{\IEEEauthorrefmark{1}Quality and Usability Lab, Technische Universität Berlin, Germany, markus.utke@campus.tu-berlin.de,\\
  saman.zadtootaghaj@qu.tu-berlin.de, steven.schmidt@tu-berlin.de}
 \IEEEauthorblockA{\IEEEauthorrefmark{2}DFKI Projektb{\"u}ro Berlin, Germany, sebastian.moeller@tu-berlin.de} 
}    

\maketitle

\begin{abstract}
\textit{} 
Video gaming streaming services are growing rapidly due to new services such as passive video streaming, e.g. Twitch.tv, and cloud gaming, e.g. Nvidia Geforce Now. In contrast to traditional video content, gaming content has special characteristics such as extremely high motion for some games, special motion patterns, synthetic content and repetitive content, which makes the state-of-the-art video and image quality metrics perform weaker for this special computer generated content. In this paper, we outline our plan to build a deep learning-based quality metric for video gaming quality assessment. In addition, we present initial results by training the network based on VMAF values as a ground truth to give some insights on how to build a metric in future. The paper describes the method that is used to choose an appropriate Convolutional Neural Network architecture. Furthermore, we estimate the size of the required subjective quality dataset which achieves a sufficiently high performance. The results show that by taking around 5k images for training of the last six modules of Xception, we can obtain a relatively high performance metric to assess the quality of distorted video games.       
\textit{}
\end{abstract}
\IEEEpeerreviewmaketitle\textit{}
\acresetall

\section{Introduction} \label{intro}
The gaming industry is one of the largest digital markets for decades which is rapidly growing with the emerging online services such as gaming video streaming, online gaming and cloud gaming (CG). While the game industry is growing, more complex games in terms of processing power are getting developed which requires players to update their end devices every few years in order to play high end games. One solution for this is to move the heavy processes such as rendering to the cloud and cut the need for high end hardware devices for customers. Cloud gaming was proposed to offer more flexibility to users in order to allow them to play any games anywhere and on any type of devices. Apart from processing power, cloud gaming benefits users by the platform independency and for game developers offers security to their products and promises a new market to increase their revenue. Besides cloud gaming, passive video streaming of gameplays become popular with hundreds of millions of viewers in a year. Twitch.tv and YouTube Gaming are the two most popular services for passive video gaming streaming. 

Quality assessment is a necessary process of any service provider to ensure the satisfaction of customers. While subjective tests are the basis of any quality assessment of multimedia services, service providers are seeking objective methods for predicting the quality, as subjective experiments are expensive and time-consuming. Depending on the amount of access to the reference signal, signal-based video quality models can be divided into three classes, no-reference (NR), reduced-reference (RR), and full-reference (FR) metrics. For QoE assessment of cloud gaming and passive video streaming services such as Twitch.tv, NR metrics are of interest for service providers as the reference signal is not available or it comes with a high cost of recording and syncing the reference signal with the distorted signal (e.g. cloud gaming). 

In this paper, we aim at designing a Convolutional Neural Network (CNN) based NR video quality metric that can predict the quality of video games with high accuracy. The main idea of our work is to train a CNN on a huge number of frames which are annotated based on a full-reference quality metric, VMAF \cite{NetflixVMAF_Github}, and then retrain a few last layers of the pre-trained CNN based on a smaller subjective image dataset. In this paper, we try to answer the following research questions before training such a huge network in the future to build the final metric:
\begin{itemize}
    \item Are machine learning based quality assessment methods suitable for computer generated imagery?
	\item How does a pre-trained deep CNN has to be retrained to gain a decent result?
    \item Which pre-trained CNN architecture performs the best among state of the art models for CGI quality assessment?
    \item How much data, in terms of number of frames, is roughly required for transfer learning of CNNs?
\end{itemize}

In order to answer these research questions without training a whole network and conducting subjective experiments blindly, pre-trained CNN architectures are taken into consideration and VMAF was chosen as a ground truth to get some insight on the selection of an architecture, the rough number of required frames for transfer learning and the expected performance.

\section{Related Work} \label{work} 
Within the last decades, we have been witness to a huge number of research works with respect to objective image and video quality assessment. In this section, due to limited space, we give a short overview of deep learning based quality models as well as metrics that are developed specifically for computer generated content. 

The performance of state of the art video and image quality metrics on gaming videos were investigated in \cite{barman2018evaluation} which shows a high correlation of VMAF with Mean Opinion Scores (MOS) while most of the NR metrics perform quite poorly. With respect to gaming content, to the best knowledge of the authors, only two NR metrics are developed. Zadtootaghaj et al. proposed a NR machine learning-based video quality metric for gaming content, named NR-GVQM, that is trained based on low level image features with the aim at predicting VMAF without having access to a reference video \cite{zadtootaghaj2018nr}. Another NR pixel-based video quality metric for gaming QoE was proposed by Goering et al \cite{GamingNR}, called Nofu. Nofu is also a machine learning metric that extracts low level features of which some are hand-crafted by the authors and some are taken from the state of the art. Nofu as a NR metric has a slightly higher performance compared to VMAF as FR metric on the GamingVideoSET \cite{barman2018gamingvideoset}.

Most of deep neural networks (DNN) based models are proposed for image quality assessment for two reasons. First, the video quality datasets are relatively small in terms of number of annotated data compared to image quality datasets due to expensive quality assessment of videos compared to image content. In addition, most of the state of the art works used transfer learning methods which uses pre-trained DNN and retrains a few last layers of it. Transfer learning is more suitable for image application as more pre-trained models are available.        
Bosse et al. \cite{bosse2018deep} presented a neural network-based approach to build FR and NR image quality metrics which are inspired by VGG16 by increasing the depth of the CNN with ten convolutional layers. 
Goering et al. [ref] proposed a hybrid NR image quality metric which is developed after training and testing a few pre-trained CNNs and extending them by adding signal-based features from state of the art models.  

\section{Datasets} \label{DS}
With the aim at training and testing the model, GamingVideoSET was used \cite{barman2018gamingvideoset}. GamingVideoSET consists of 24 source video sequences from 12 games (two sequences per game), which are encoded using H.264/MPEG-AVC under different bitrate-resolution pairs. 18 video sequences from 10 games were selected and used in the training process of the model, while six source video sequences from four games were selected for the validation set. With the aim to investigate the suitability of ML methods for gaming content because of similarity between video sequences from the same game, two video sequences in the validation set were chosen from the same games that are included in the training set (but different recorded sequences), and four other video sequences were selected from games that are not in the training set.  
In total, we selected among \SI{279900} frames in the training set and \SI{71100} frames in the validation set as explained later.

\section{Experiments and Results} \label{Results}
To test if deep CNN are applicable on our type of data we use \textit{transfer learning}, denoting the process of retraining or fine tuning a pre-trained neural network to make it perform a task that is different from the one it was originally trained to do. Using Keras\cite{chollet2015keras}, a high level neural network library, different model architectures are available along with their pre-trained weights on the ImageNet\cite{russakovsky2015imagenet} database. We chose three different architectures and compared their performance on our dataset in table \ref{tab:architectures}: DenseNet121\cite{huang2017densely}, ResNet50\cite{he2016deep}, and Xception\cite{chollet2017xception}. For each of these architectures the fully connected network at the end was removed. Instead we used one dense layer consisting of only one output neuron with linear activation. The output of the network is directly compared to the actual VMAF value. To calculate the quality at the video level, the frame level values are averaged.

Because of the large size of the images in the dataset we cannot efficiently train the network on the images directly. Hence, we crop random patches of size $299\times299$ from the frames we want to train on. It has to be noted that this size is the standard input size of Xception. This is done in parallel to the training, such that in each epoch a new random patch of each image is chosen.

We used the Xception architecture for the following investigations since it worked best for our setting (see table \ref{tab:architectures}). However, it should be said that for an extensive evaluation multiple parameter settings should have been compared.

\begin{table}[h]
    \centering
    \begin{tabular}{
            @{} 
            cc @{\qquad}
            S[table-format=2.3] @{\quad}
            S[table-format=2.3] @{\quad}
            S[table-format=2.3] @{\quad}
            S[table-format=2.3] @{}
            }
        \toprule
        Architecture &Level&\multicolumn{1}{c}{R$^2$} \quad & \multicolumn{1}{c}{RMSE} \quad & \multicolumn{1}{c}{PCC} \quad & \multicolumn{1}{c}{SRCC}\\
        \midrule
        \multirow{2}{*}{DenseNet121} & frame  & 0.87 & 7.43 & 0.934 & 0.937 \\
        & video  & 0.97 & 3.38 & 0.985 & 0.984 \\[5pt]
        \multirow{2}{*}{ResNet50} & frame  & 0.87 & 6.98 & 0.938 & 0.939 \\
        & video  & 0.97 & 3.11 & 0.987 & 0.985 \\[5pt]
        \multirow{2}{*}{Xception} & frame  & 0.87 & 6.85 & 0.939 & 0.940 \\
        & video  & 0.98 & 2.75 & 0.990 & 0.988 \\
        \bottomrule
    \end{tabular}
    \caption{Results for different architectures}
    \label{tab:architectures}
\end{table}

\begin{figure*}
    \centering
    \begin{subfigure}{0.32\textwidth}
        \centering
        \includegraphics[width = \textwidth]{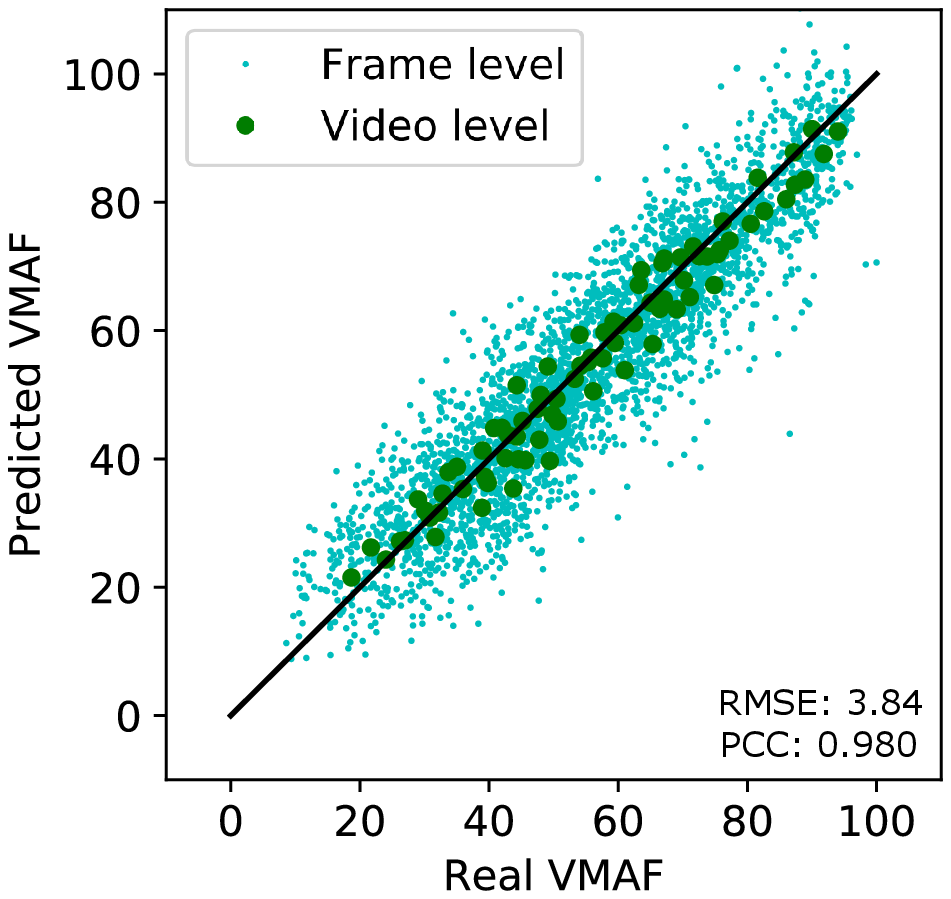}
        \caption{One module}
    \end{subfigure}
    \begin{subfigure}{0.32\textwidth}
        \centering
        \includegraphics[width = \textwidth]{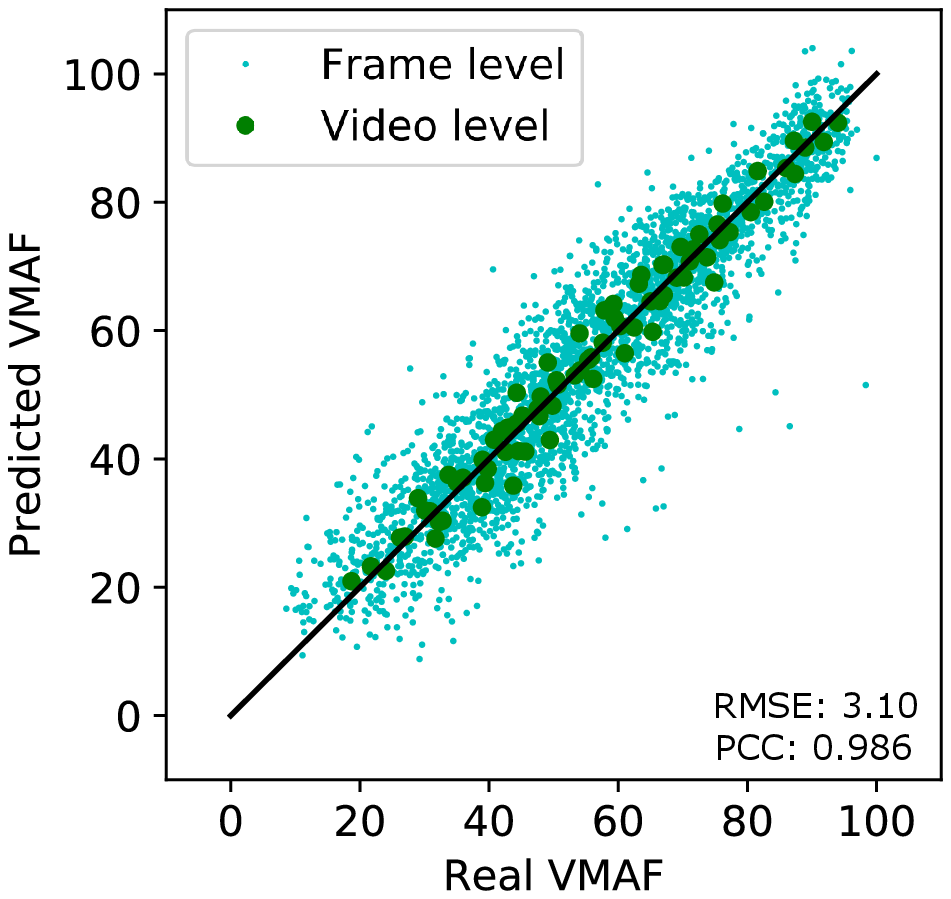}
        \caption{Four modules}
    \end{subfigure}
    \begin{subfigure}{0.32\textwidth}
        \centering
        \includegraphics[width = \textwidth]{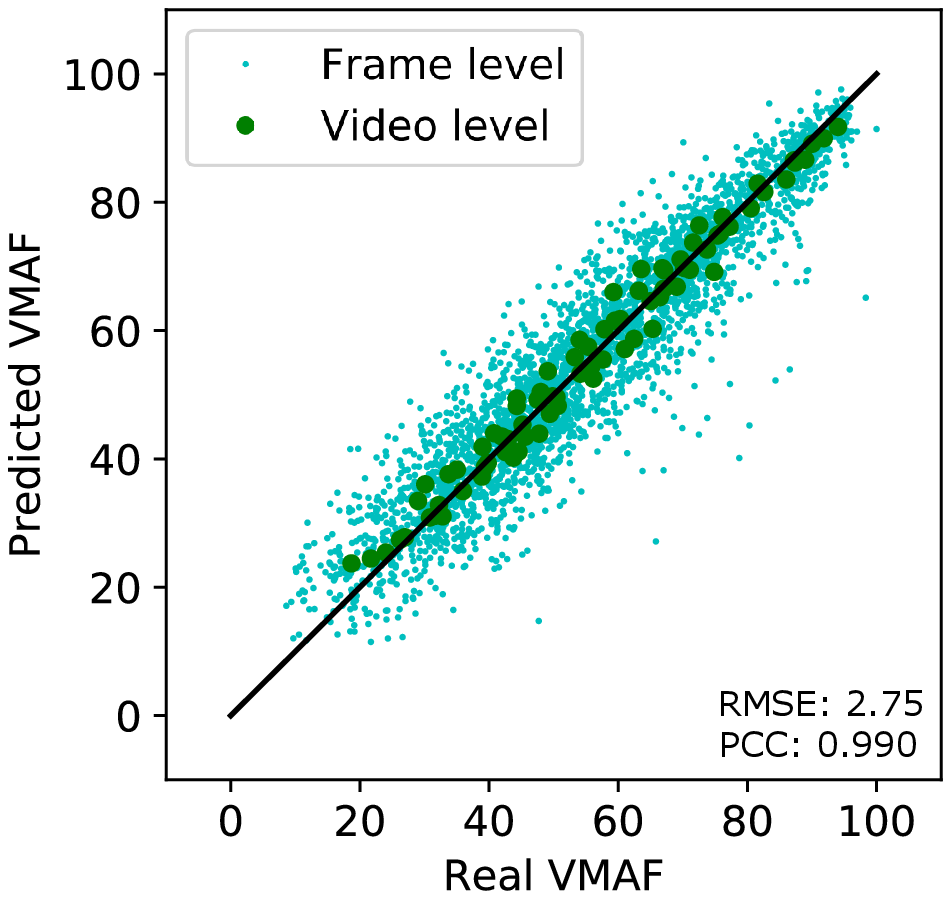}
        \caption{Six modules}
    \end{subfigure}
    \caption{Different number of trained modules of the Xception network (RMSE and PCC on video level)}
    \label{fig:nLayers}
\end{figure*}

When using pre-trained networks it is often sufficient to train only the last layers instead of retraining the whole network. To investigate this we tried training different amounts of the network on our dataset and compared the performance. The Xception architecture consists of 14 modules, each containing two to three convolutional layers. Figure \ref{fig:nLayers} shows a comparison of the results when training only the last, last four and last six modules. We observe that the results get better when more modules are used in the training process. However, this might increase the risk of overfitting as the dataset is not large enough. If the performance can get improved even more by training more than six modules is yet to be tested which may come with higher cost of computation.

From the 12 different games in the dataset two are used only in the validation set and two are present in both validation and training set. Surprisingly comparing the results for these two groups showed no difference. 

In the future we want to train our model on subjective ratings instead of VMAF values. Therefore, we tried to find a small subset of our data, that still performs well. 
Since the training dataset consists of over \SI{25000} frames and consecutive frames can be very similar, only every $n$-th frame from every video is used for training. A very high number for $n$ was chosen ($n = 403$) and then lowered step by step to find the point, where the performance of the model stops improving. Table \ref{tab:numberOfSamples} shows the RMSE and SRCC for different choices of $n$. To provide good results and minimize the number of frames needed, $n = 53$ seems to be a good choice.

We also investigated if only cropping the center of images would lead us to better performance compared to randomized cropping during the training of the CNN (as users tend to look more at the center of images). Our results showed that it cannot improve the network and taking random patches for training would gain much higher performance.  

\begin{table}
    \centering
    \begin{tabular}{
            @{} 
            S[table-format=3] 
            S[table-format=4]
            S[table-format=1.2] 
            S[table-format=1.2] @{\qquad}
            S[table-format=1.3] 
            S[table-format=1.3] @{}
            }
        \toprule
         \multicolumn{1}{c}{\multirow{2}{*}{$n$}} & \multicolumn{1}{c}{\multirow{2}{*}{\shortstack{Total number\\ of frames}}} & \multicolumn{2}{c}{RMSE} \qquad & \multicolumn{2}{c}{SRCC} \\
         &&\multicolumn{1}{c}{frame}&\multicolumn{1}{c}{video} \qquad&\multicolumn{1}{c}{frame}&\multicolumn{1}{c}{video} \\
        \midrule
        403 & 933 & 7.50 & 4.40 & 0.939 & 0.981\\
        203 & 1555 & 7.12& 4.07 & 0.938 & 0.980 \\
        103 & 2799 & 7.18& 3.72 & 0.940 & 0.985 \\
        53  & 5287 & 6.78& 3.11 & 0.942 & 0.987 \\
        23  & 9952 & 7.09& 3.19 & 0.937 & 0.986 \\
        \bottomrule
    \end{tabular}
    \caption{RMSE and SRCC for different choices of $n$}
    \label{tab:numberOfSamples}
\end{table}

\section{Discussion and Conclusion} \label{Dis}
This paper is presented to demonstrate the effectiveness of the usage of CNNs for quality assessment of multimedia services. While such CNN-based quality metrics come with high computation cost which might not be suitable specially for real-time services such as cloud gaming, the high performance of these methods motivate us to consider them as a future of QoE assessment for some special use cases such as quality assessment of uploaded video content for transcoding purposes, e.g. for Twitch.tv. It has to be noted that this paper only presents the performance of a few examples of pre-trained CNNs that aim to predict a FR metric, VMAF. However the authors expect to reach similar performance when real MOS values are used as a ground truth. The initial results show that with a medium sized image quality dataset, between 1k and 5k annotated images, the quality can be predicted with relatively high performance. It was observed that with 5k annotated frames, retraining six modules of Xception might avoid over fitting while also getting high performance. Our plan is to reduce this number significantly by mixing the training of CNNs with VMAF and MOS values, which requires a smaller dataset of annotated frames by MOS values. In addition, we try to use sophisticated techniques to pool the frame-level quality values to obtain video quality such as using Long Short-Term Memory (LSTM) architecture.   

Moreover, we observed that using videos from the games that are used in the training and validation set did not lead us to significantly higher performance.  However, we need to investigate this more as we only tested it with a small dataset.
Another observation was that making the network deeper might help in order to get better performance as it was also observed in \cite{bosse2018deep}. However, that depends on the size of the training dataset as well as on its diversity of content. Otherwise, training a deep CNN could lead to overfitting.

\bibliographystyle{ieeetr}

\end{document}